\documentclass[twocolumn,a4paper,showpacs,preprintnumbers,amsmath,amssymb,nofootinbib,floatfix]{revtex4}
\input psfig.sty  
\def \be {\begin{equation}} 

\def \ee {\end{equation}} 
\def \bea {\begin{eqnarray}} 
\def \eea {\end{eqnarray}} 

\usepackage{graphicx}
\usepackage{dcolumn}
\usepackage{bm}
\usepackage{epsfig} 

\begin{document}
\title{Searching for cosmological signatures of the Einstein equivalence principle breaking}
\author{R. F. L. Holanda$^{1,2}$  } \email{holanda@uepb.edu.br}
\author{K. N. N. O. Barros$^3$}\email{kleberbarros@cct.uepb.edu.br}
\affiliation{$^1$ Departamento de F\'{\i}sica, Universidade Estadual da Para\'{\i}ba, 58429-500, Campina Grande - PB, Brasil,
\\$^2$ Departamento de F\'{\i}sica, Universidade Federal de Campina Grande, 58429-900, Campina Grande - PB, Brasil,\\
$^3$ Departamento de Estat\'{\i}stica, Universidade Estadual da Para\'{\i}ba, 58429-500, Campina Grande - PB, Brasil.}

\date{\today}

\begin{abstract}

 Modifications of gravity generated by a multiplicative coupling of a scalar field to the electromagnetic Lagrangian lead to a breaking of Einstein equivalence principle (EEPB) as well as to variations of fundamental constants. In these theoretical frameworks, deviations of standard values of the fine structure  constant, $\Delta \alpha/\alpha=\phi$, and of the cosmic distance duality relation, $D_L(1+z)^{-2}/D_A=\eta=1$, where $D_L$ and $D_A$ are the luminosity and angular diameter  distances, respectively,  are unequivocally linked. In this paper, we search for cosmological signatures of the EEPB by using angular diameter distance  from galaxy clusters, obtained via their Sunyaev-Zeldovich effect (SZE) and X-ray observations, and distance modulus of type Ia supernovae (SNe Ia). The crucial point here is that we take into account the dependence of the SZE/X-ray technique with $\phi$ and $\eta$.  Our new results show no  indication of the EEPB.

\end{abstract}
\pacs{98.80.-k, 95.36.+x, 98.80.Es}
\maketitle

\section{Introduction}

The so-called reciprocity relation, proved long ago by Etherington (1933, 2007), is a fundamental result for observational
cosmology and its most useful version in the astronomical context, known as cosmic distance duality relation (CDDR),  is defined by
\begin{equation}
\frac{D_L}{D_A}(1+z)^{-2}=\eta=1
\label{eq1}
\end{equation}
which relates the luminosity distance $D_L$ with the angular diameter distance $D_A$  on the same redshift $z$. This equation is completely general, only requires that source and observer are connected by null geodesics in a Riemannian spacetime and that the number of photons is conserved (Ellis, 2007). 
\begin{figure*}
\centering
\includegraphics[width=0.47\textwidth]{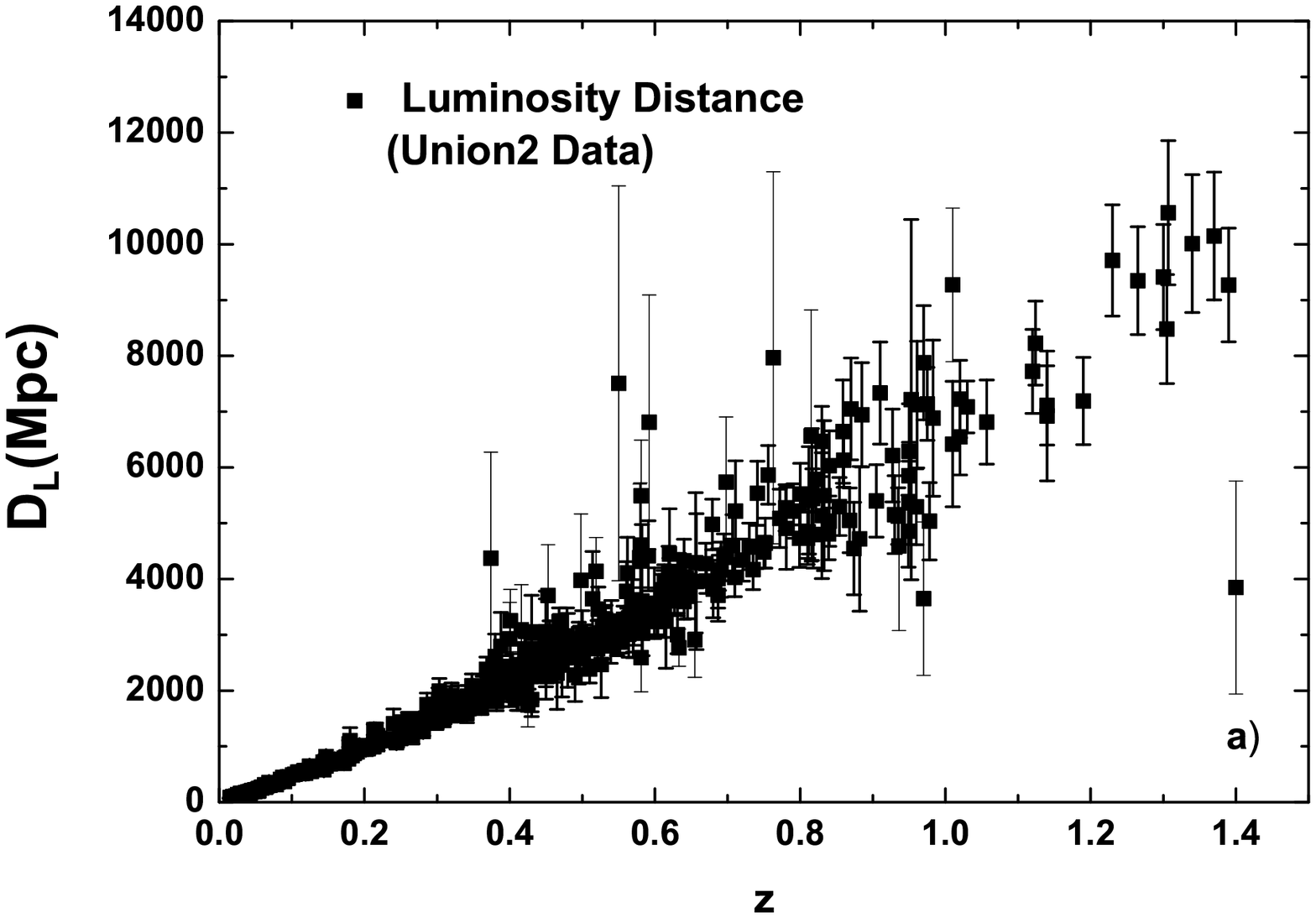}
\hspace{0.3cm}
\includegraphics[width=0.47\textwidth]{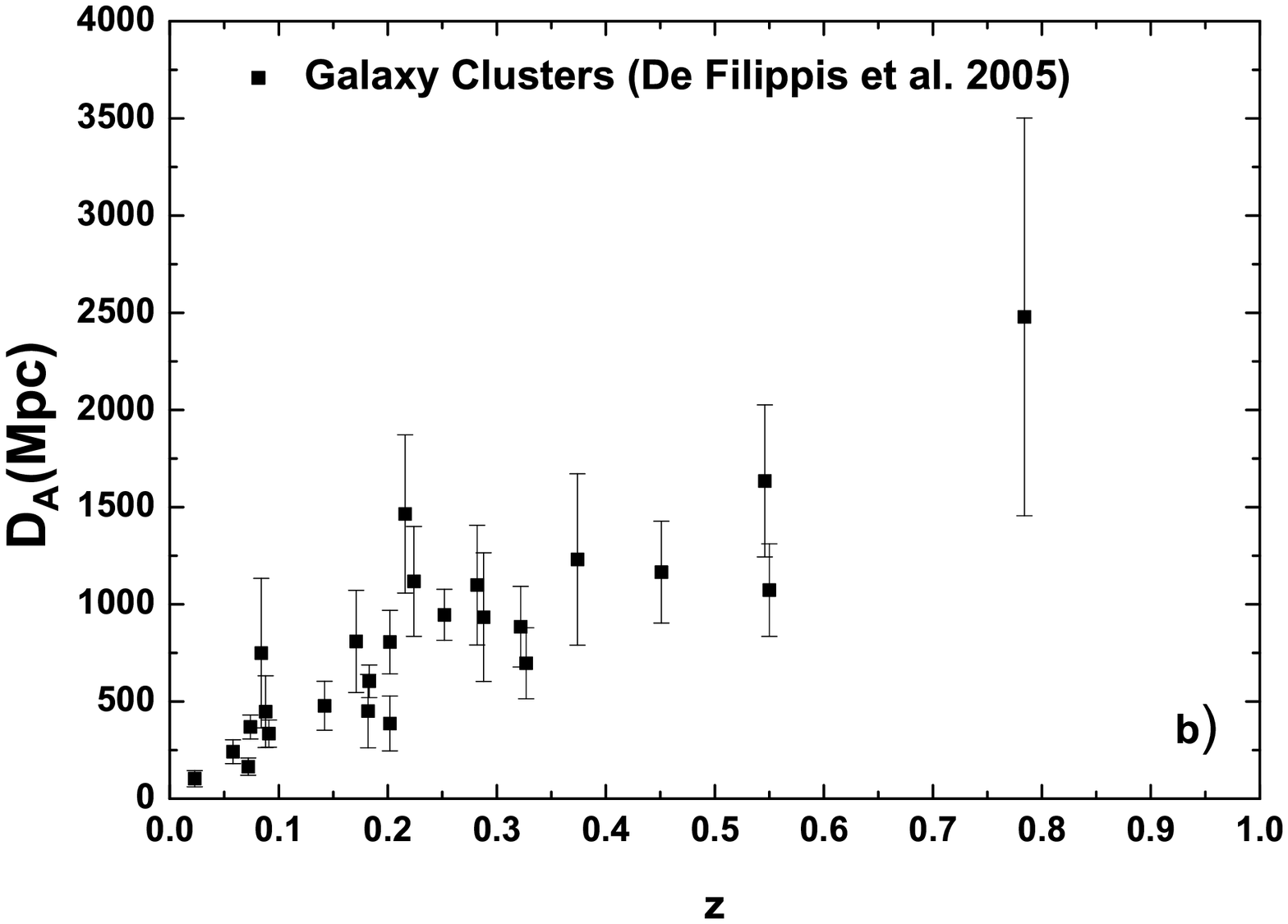}
\caption{(a) Luminosity distance of the Union2 SNe Ia Sample as a function of redshift. (b) Angular diameter distance measurements as a function of redshift  from De Filippis et al. (2005) sample. The red points are the luminosity distances in each galaxy cluster obtained by taking  the weighted average of the SNe Ia observational quantities.
}
\end{figure*}
Recently, Hees,  Minazzoli \& Larena (2014) investigated cosmological signatures of modifications of gravity generated by a multiplicative coupling of a scalar field to the electromagnetic Lagrangian. These authors showed, via the geometric optics approximation, that in these frameworks photons propagate on null geodesics, but their number is not conserved, violating the Eq. (1). Moreover, this kind of coupling  
leads to variations of the fine structure constant $\alpha$, $\Delta \alpha/\alpha = \phi$, that is  intimately and unequivocally linked to violations of the CDDR, $\eta \neq 1$, by $\phi(z)=\eta^2(z)$. In this context, there are many hypothetical alternative theories of gravity such as: the low energy action of string theories (Damour \&  Polyakov 1994; Gasperini, Piazza \&  Veneziano 2001; Minazolli 2014), in the context of axions (Peccei \&  Quinn (1977); Dine,  Fischler \& Srednicki (1981); Kaplan (1985)), of generalized chameleons (Brax et al. 2004;  Brax, van de Bruck \&  Davis 2007; Ahlers et al. 2008), in Kaluza-Klein theories with additional compactified dimensions (Overduin \&  Wesson 1997; Fujii \& Maeda 2003), in the Bekentein-Sandvik-Barrow-Magueijo theory of varying $\alpha$ ( Bekenstein 1992; Sandvik, Barrow \& Magueijo 2002; Barrow \& Lip 2012; Barrow \& Graham 2013), in extended $f(R,L_m)$ gravity (Harko, Lobo \&  Minazzoli 2013) or in the context of the pressuron theory (Minazzoli \& Hees 2013).  This class of extensions of the general relativity leads to an  Einstein Equivalence Principle breaking (EEPB) in the electromagnetic sector. 

By using different astronomical quantities,  several papers have tested the CDDR relation in the last years: SNe Ia plus $H(z)$ data (Avgoustidis et al. 2009, 2010; Holanda, Carvalho \& Alcaniz 2013; Liao, Avgoustidis \& Li 2015), gas mass fractions of galaxy clusters and SNe Ia (Gon\c{c}alves, Holanda \& Alcaniz 2012; Holanda, Gon\c{c}alves \& Alcaniz 2012), gamma-ray burst plus $H(z)$ (Holanda \& Busti 2014), SNe Ia plus barion acoustic oscillations (BAO) (Puxun et al. 2015), cosmic microwave background radiation (Ellis et al. 2013), gas mass fraction plus H(z) data (Santos-da-Costa, Busti \& Holanda 2015), SNe Ia plus cosmic microwave background (CMB) and BAO (Lazkoz et al. 2008), gravitational lensing plus SNe Ia (Holanda, Busti \& Alcaniz 2016; Liao et al. 2015). As a result, the validity of the CDDR was verified at least within 2$\sigma$  (see table in Holanda, Busti \& Alcaniz 2016).

On the other hand, some other tests used the angular diameter distance (ADD) of galaxy clusters via Sunyaev-Zeldovich effect (SZE) (Sunyaev \& Zel’dovich 1972) and X-ray observations (see, for instance, Uzan et al. 2004, Holanda, Lima \& Ribeiro 2010; Holanda, Lima \& Ribeiro 2012; Li et al. 2011; Nair,  Sanjay \& Deepak 2011; Yang et al. 2013). However, it was shown very recently that the SZE/X-ray technique for measuring ADD of galaxy clusters depends on not only  of the CDDR validity, but also on the fine structure constant, $\alpha$ (Holanda et al.  2016a, 2016b). If $\alpha(z)=\alpha_0 \phi(z)$, where $\alpha_0$ is the current value, the SZE and X-ray measurements do not give the true ADD  but $D_A^{data}=\phi(z)\eta^2D_A$ (other papers considered only $D_A^{data}=\eta^2D_A$). In this way, in the framework proposed by Hees,  Minazzoli \& Larena (2014), the tests of CDDR involving the SZE/X-ray technique have questionable estimates of $\eta(z)$. 

In this paper, by considering the large class of theories proposed  by Hees,  Minazzoli \& Larena (2014), we search for deviations of EEP by putting limits on the $\eta(z)$ parameter via SNe Ia and ADD of galaxy clusters obtained via the SZE/X-ray technique by  taking into account the dependence of the SZE/X-ray technique with $\phi$ and $\eta$. {  In order to test the Eq. (1)  it is necessary SNe Ia and galaxy clusters with identical redshifts. In this way, we consider the SNe Ia Union2 compilation (Amanullah et al. 2010) and the 25 ADD of galaxy clusters compiled by De Filippis et al. (2005) as follows: for each galaxy cluster, we select  SNe Ia  with redshifts obeying the  criteria  $|z_{cluster} - z_{SNe}| \leq 0.005$, resulting in 25 SNe Ia sub-samples which matched this criterion. Then,  we take the weighted average  of the SNe Ia observational quantities in each sub-sample to perform our analyses. Moreover, since the Union2 compilation provides  the distance modulus  of  SNe Ia, which has a  dependence on a cosmological model, and their original and cosmological model-independent data,  we perform our analyses by using these two quantities.}

The paper is organized as follows. In Section 2, we briefly describe the SNe Ia
data (Amanullah et al. 2010) and the ADD data (De Filippis et al. 2005). In Section 3, we describe the methods used in our analyses. In section 4, we perform the analyses. Finally, the discussions and conclusions are given in Section 5.


\section{Samples}
\label{sec:test3}

- The full SNe Ia sample is formed by 557 SNe Ia data compiled by Amanullah et al. (2010), the so-called Union2 compilation. Besides adding 250 SNe Ia in the Union compilation (Kowalsi et al. 2008), Amanullah and co-workes (2010) refined the Union analysis refitting  all light curves with the SALT2 fitter (Guy et al. 2007). However, in this fitter the distance modulus of SNe Ia sample depend on cosmological model (usually $\omega$CDM) and the Hubble constant. So, we consider  two methods in our analyses:   Method (I) we use the cosmological model-dependent distance modulus  of SNe Ia and  method (II) we use  their $\mu(\alpha, \beta,M_B)=m^{\rm max}_{B}-M_{B}+\alpha x-\beta c$ measurements, where $m^{max}_B$ is the rest-frame peak magnitude of B bands, $x$ is stretch factor,  describing the effects of shapes of light curves, and $c$ is color parameter, representing the influences of the intrinsic color and reddening by dust on $\mu(\alpha, \beta,M_B)$. These three parameters are cosmological model independent since they are obtained by fitting the light curves of SNe Ia. 

- {  The second sample is formed  by the 25 ADD of galaxy clusters from the De Filippis et al.  (2005). Motivated by the images from Chandra and XMM-Newton telescopes, which show a elliptical surface brightness of galaxy clusters, these authors used an isothermal elliptical $\beta$ model to describe the galaxy clusters. The ADD were derived for two sub-samples discussed in the literature where a spherical $\beta$ model was assumed. The first one, compiled by Reese et al. (2002), is a selection of 18 galaxy clusters.The second sub-sample of Mason et al. (2001) has 7 clusters from the X-ray limited flux sample of Ebeling et al. (1996). The galaxy clusters are distributed over the redshift interval $0.023 \leq z \leq 0.784$. It is worth mentioning that in the Hees,  Minazzoli \& Larena (2014) framework, the SZE observations are also affected by variations of $\alpha$ and $\eta$ due to a change in the law of evolution of the CMB, such as, $T_{CMB}(z)=T_0(1+z)^{1+\varsigma}$, where  $T_0$ is the current value. However, the frequency used to obtain the SZE signal in galaxy clusters sample of De Filippis et al. (2005) was 30 GHz, in this band the effect on the SZE from a variation of $T_{CMB}$ is completely negligible (F. Melchiorri and O. Melchiorri 2005; Saro et al. 2014). 

In Fig. (1a) we plot the Union2 sample. In Fig. (1b) we plot the De Filippis et al. (2005) sample.}\\

\section{Samples}
\label{sec:test1}
\begin{figure*}
\centering
\includegraphics[width=0.47\textwidth]{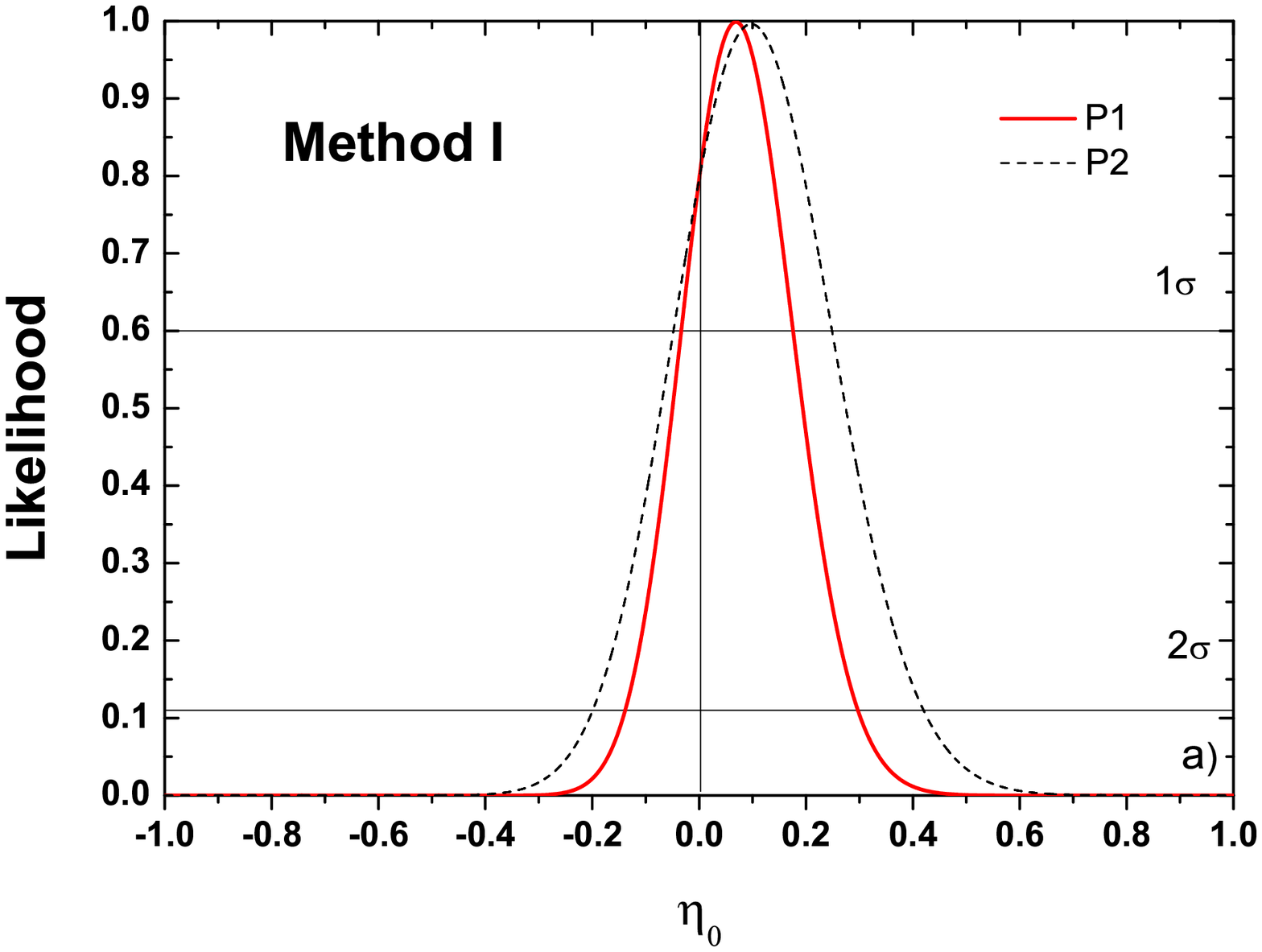}
\hspace{0.3cm}
\includegraphics[width=0.47\textwidth]{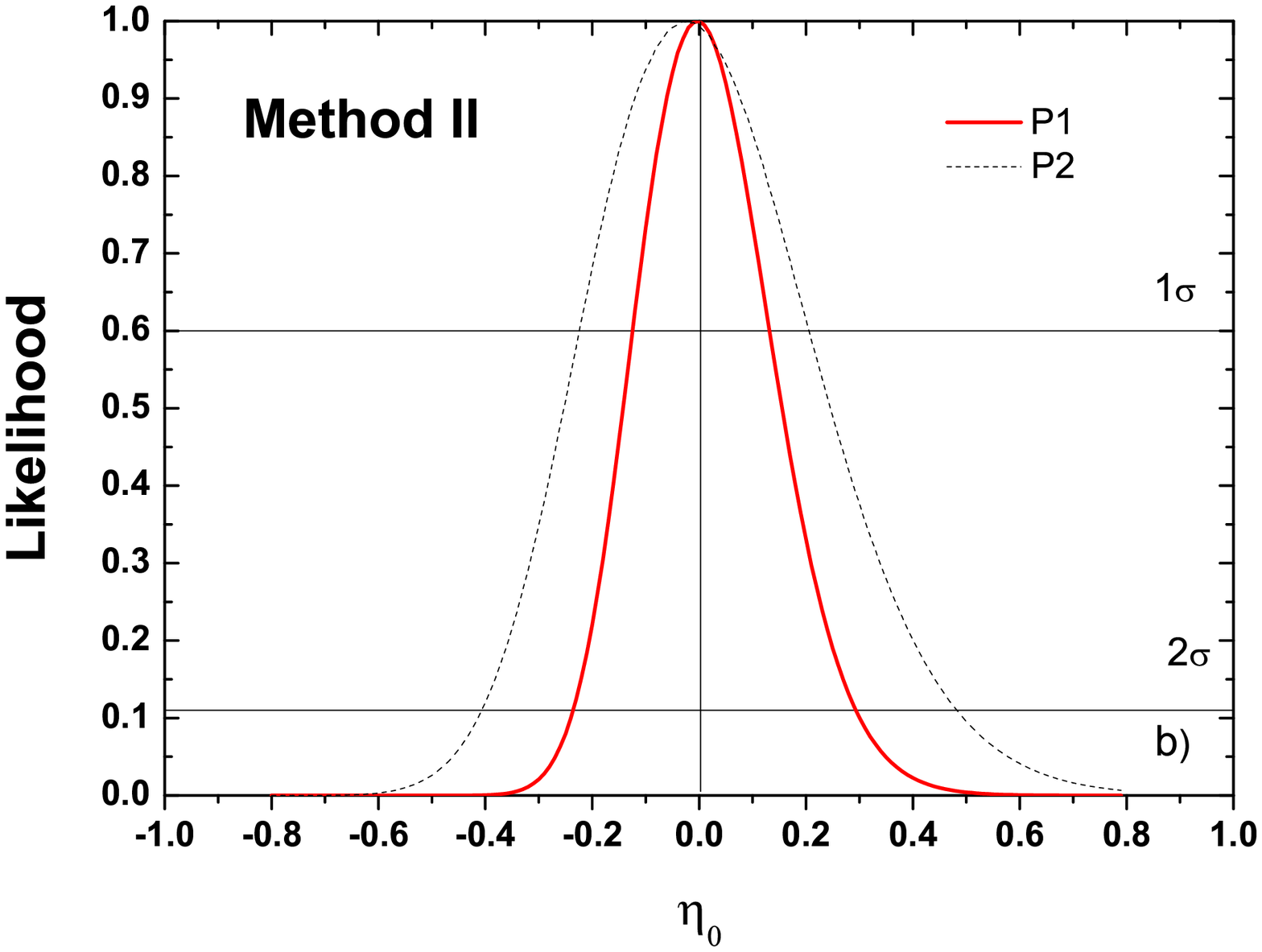}
\caption{Likelihoods of the $\eta_0$ parameter. In figures (a) and (b) we display the results from method (I) and (II). In both figures, the red solid lines and the dashed black lines are for the functions P1 and P2, respectively. In our analyses, we have added in quadrature the statistical and systematic errors in galaxy clusters data. Following  Amanullah et al. (2010) we added  a 0.15 systematic error to SNe Ia data.
}
\end{figure*}

\subsection{Methods and Analyses}

{ The estimates of the $\eta_0$ parameter are obtained from two methods, namely:\\

- Method I:  we consider the 25 ADD of galaxy clusters from De Filippis et al. (2005) sample and,  for each i-galaxy cluster, we obtain one  distance modulus, $\bar{\mu}$, and its error, $\sigma^2_{\bar{\mu}}$, from all i-SNe Ia  with $|z_{cluster_i} - z_{SNe_i}| \leq 0.005$\footnote{ { This criterion allows us to have some SNe Ia for each galaxy cluster and so we can perform  a  weighted average with them in order to minimize the scatter observed on the Hubble diagram (see fig. 1a). Moreover, if one considers that we live in an universe close to the cosmic concordance model, the  criteria $\Delta z \leq 0.005$ implies an error on the distances lower than 5\% for $z \geq 0.1$, therefore, the bias on our results from this criterion are contemplated in galaxy clusters errors.}} by calculating the following weighted average (Meng et al. 2012):

\begin{equation}
\begin{array}{l}
\bar{\mu}=\frac{\sum\left(\mu_{i}/\sigma^2_{\mu_{i}}\right)}{\sum1/\sigma^2_{\mu_{i}}} ,\\
\sigma^2_{\bar{\mu}}=\frac{1}{\sum1/\sigma^2_{\mu_{i}}}.
\end{array}\label{eq:dlsigdl}
\end{equation}
 }

On the other hand, as discussed by Holanda et al. (2016b), for modifications of gravity via the presence of a scalar field with a multiplicative coupling to the electromagnetic Lagrangian, the SZE and X-ray measurements do not give the true distance, but $D_A^{data}=\eta^4(z)D_A$.  Moreover, as argued by Holanda, Lima \& Ribeiro (2010), if one wants test the CDDR  by using $D_L(1+z)^{-2}D_A^{-1}=\eta$ and galaxy clusters observations, the angular diameter distance $D_A(z)$ must be replaced by $D_A(z)=\eta^{-4}D_A^{data}$ ($\eta^{-2}$ in their case, since variations of $\alpha$ were not considered). In this way, we have access to 

\begin{equation}
\frac{D_L}{(1+z)^2D_A^{data}(z)}=\eta^{-3}(z).
\end{equation}
By using the equation above, we define the distance modulus of a galaxy cluster data as

\begin{equation}
 \mu_{cluster}(\eta ,z)=5\lg[\eta^{-3}(z)D^{data}_A(z)(1+z)^2]+25. 
\end{equation}

We evaluate our statistical analysis by defining the likelihood distribution function, $L \alpha \exp(-\chi^{-2}/2)$, where

\begin{eqnarray}
\chi^2=\sum_{i=1}^{25}\frac{(\bar{\mu}(z_i)-\mu_{cluster}(\eta,z_i))^2}{\sigma_{obs}^2},
\label{chi}
\end{eqnarray} 
with $\sigma_{obs}^2= \sigma^2_{\bar{\mu}} + \sigma^2_{\mu cluster}$. The sources of statistical uncertainty in the error bars of $D_{A}^{\: data}$(z)  are:  SZE point sources $\pm 8$\%, X-ray background $\pm 2$\%, Galactic N$_{H}$ $\leq \pm 1\%$, $\pm 8$\% kinetic SZ and for CMB anisotropy $\leq \pm 2\%$. We have added in quadrature the following systematic errors: SZ calibration $\pm 8$\%, X-ray flux calibration $\pm 5$\%, radio halos $+3$\% and X-ray temperature calibration $\pm 7.5$\%. Following  Amanullah et al. (2010) we added  a 0.15 systematic error to SNe Ia data.

- Method II: consists of using the very same SNe Ia of the method I, but now considering their raw data, i.e., their $m^{\rm max}_{B}$, $x$ and $c$ measurements from the function $\mu(\alpha, \beta,M_B)=m^{\rm max}_{B}-M_{B}+\alpha x-\beta c$. In this way, by using the Eq. (2), we obtain: $\bar{\mu}(\alpha, \beta,M_B)=\bar{m}^{\rm max}_{B}-M_{B}+\alpha \bar{x}-\beta \bar {c}$. Here, our main aim is to explore how much the cosmological model adopted in distance modulus influences the results.  For this case, the likelihood estimator is written as

\begin{align}
&\chi^2(\alpha, \beta, M_B, \eta) \nonumber \\
& =\sum_{i=1}^{25}\frac{\big[\bar{\mu}(\alpha,\beta,M_B;z_i)-\mu_{\rm cluster}(\eta;z_i)\big]^2}
{\sigma^2_{\rm total}(z_i)}, \label{newchi2}
\end{align}
where  the uncertainty $\sigma^2_{\rm total}(z_i)$ is given by
\begin{equation}
\sigma^2_{\rm total}(z)=\sigma^2_{\bar{m}}(z)+\alpha^2\sigma^2_{\bar{x}}(z)+\beta^2\sigma^2_{\bar{c}}(z)
+\Big[\frac{5}{\ln10}\cdot\frac{\delta_{D_A^{data}}(z)}{D_A^{data}}\Big]^2,
\end{equation}
where  $\sigma_{\bar{m}}$,  $\sigma_{\bar{x}}$, $\sigma_{\bar{c}}$, and $\delta_{D_A}$ are the errors of $\bar{m}^{\rm max}_{B}$, $\bar{x}$, $\bar{c}$, and $D^{\rm cluster}_A$ respectively. We adopt an iterative  method to calculate the likelihood for $\eta_0$ with step size $0.01$ for all parameters. We use the following flat priors for the parameters: $\eta_0 = [-2.0, 2.0]$, $\alpha =[-0.6, 0.6]$,  $\beta = U[-1.0, 10.0]$ and $M_B = U[-21, -18]$.

In both methods, in order to search deviations of EEP we consider the following functions to $\eta(z)$ (Holanda, Lima \& Ribeiro 2010): $\eta(z)=1+\eta_0 z$ (P1) and $\eta(z)=1+\eta_0 z/(1+z)$ (P2). The first expression is a continuous and smooth one-parameter linear expansion, whereas the second one includes a possible epoch-dependent correction, which avoids the divergence at extremely high-z. {  Naturally, these functions deforming the CDDR retrieve the equality between the distances for $z = 0$, 
since in this case there is no cosmic expansion, i. e., the space is Euclidean. Moreover, as one may see, if the likelihood of $\eta_0$ to peak at $\eta_0 = 0$ the EEP is satisfied. }

{  The results from the method I  are plotted in Fig. (2a). The red solid line and the dashed black line are from the functions P1 and P2, respectively. The constraints on $\eta_0$ parameter are, respectively: $\eta_0=0.069 \pm 0.106$ and $\eta_0=0.097 \pm 0.152$  at 1$\sigma$ c.l.. As one may see the results are completely equivalent  and they show no violation of the CDDR or, equivalently, no indication of EEPB.

 We  plot in Fig. (2b) the results from the method II. Again, the red solid line and the dashed black line are from the functions P1 and P2, respectively. The constraints on $\eta_0$ parameter are, respectively: $\eta_0=-0.00 \pm 0.135$ and $\eta_0=-0.03 \pm 0.20$  at 1$\sigma$ c.l.. Interestingly , the likelihoods from Method I are slightly displaced to the right, but still compatible with $\eta_0 =0$ at 1 $\sigma$ c.l..

We can compare our results with previous works that used the De Filippis et al. (2005)  and Union2 samples to test the CDDR regardless possible variations on $\alpha$. For instance, for (P1) and (P2) functions, respectively: Li, Wu \& Yu (2011) found $\eta_0=-0.07 \pm 0.19$ and$\eta_0=-0.11 \pm 0.26$;  Meng et al. (2012) found $\eta_0=-0.047 \pm 0.178$-$\eta_0=-0.083 \pm 0.246$. These results should be compared only with those ones from our method I. As one may see, their values are less restrictive  and slightly different of our results.  } 


\section{Conclusions}

 As it shown recently, modifications of gravity occurring via the presence of a scalar field with a multiplicative coupling to the electromagnetic Lagrangian lead to an explicit Einstein Equivalence Principle breaking (EEPB). In this theoretical framework,  variations of the fine structure constant, $\phi = \Delta \alpha/\alpha$, and violations of the distance duality, $\eta \neq 1$ (see Eq. 1), are unequivocally linked by $\phi(z)=\eta^2(z)$ (Hees, Minazzoli \& Lorena 2014). In this paper, we have shown that  angular diameter distance of galaxy clusters based on their Sunyaev-Zeldovich effect and X-ray surface brightness observations, $D^{data}_A$,  can be used to search for cosmological signature of the EEPB. As argued by Holanda et al. (2016b), current $D^{data}_A$ of galaxy clusters do not give the true distance $D_A$, but $D^{data}_A=D_A \phi(z)\eta^2(z)$. In this way, by assuming the relation $\phi(z)=\eta^2(z)$ valid in Hees, Minazzoli \& Lorena 2014 framework,  the quantity  $D^{data}_A=D_A \eta^4(z)$ was used  jointly with SNe Ia observations to search for deviations of standard values of $\alpha$ and $\eta$.

We have searched  the EEPB by using  two well-known functions to $\eta(z)$, such as, $\eta(z)=1+ \eta_0 z$ (P1) and $\eta(z)=1+ \eta_0 z/(1+z)$ (P2). The observational quantities considered were: 25 $D^{data}_A$ from De Filippis et al. (2005) and 25 SNe Ia sub-samples from Union2  compilation (Amanullah et al. 2010), where each sub-sample contains SNe Ia with $|z_{cluster} - z_{SNe}| \leq 0.005$. We have used two methods: in the method I we have considered the  weighted average of the cosmological model depend distance modulus of the SNe Ia in each sub-sample and in the method II,  the weighted average  of the the original data of the SNe Ia , i.e., their   $\mu(\alpha, \beta,M_B)=m^{\rm max}_{B}-M_{B}+\alpha x-\beta c$ measurements, which are cosmological model independent. Our results have shown  no  indication of violation of the EEP, with the likelihoods peaked close to $\eta_0=0$ at 1$\sigma$ c.l. for all methods.

\acknowledgements

RFLH acknowledges financial support from INCT-A and CNPq (No. 478524/2013-7, 303734/2014-0). The author RFLH thanks Vinicius Consolini Busti for useful comments and suggestions.

\label{lastpage}

\begin{thebibliography}{99}

\bibitem{ah}  Ahlers, M. et al., 2008, PRD, 77, 015018 
\bibitem{aman}Amanullah, R. et al., 2010, ApJ, 716, 712
\bibitem{Avgoustidis10} Avgoustidis, A., Burrage, C., Redondo, J., Verde, L. \& Jimenez, R., 
2010, JCAP, 10, 024
\bibitem{Avgoustidis12} Avgoustidis, A.,  Verde, L. \& Jimenez, R., 2009, JCAP, 0906, 012
\bibitem{barrow} Barrow, J. D. \&  Lip, S. Z. W., 2012, PRD, 85, 023514
\bibitem{barrow2}  Barrow, J. D. \&  Graham, A. A. H., 2013, PRD, 88, 103513 
\bibitem{bej}Bekenstein, J. D., 1982, PRD, 25, 1527 
\bibitem{brax} Brax, P. et al., 2004, PRD,  70, 123518
\bibitem{brax2} Brax, P., Van de Bruck, C. \&  Davis, A. C.,  2007, PRL, 99, 121103 
\bibitem{Bassett:2003vu} Bonamente, M., et al. 2006, ApJ, 647, 25
\bibitem{cav}Cavaliere, A., \& Fusco-Fermiano, R. 1978, A\&A., 667, 70
\bibitem{deber} Damour, T.\&  Polyakov, A.M., 1994, Nuclear Physics B, 423, 532
\bibitem{de}De Filippis, E., Sereno, M., Bautz, M.W., \& Longo, G. 2005, ApJ, 625, 108
\bibitem{dini} Dine, M.,  Fischler, W., \&  Srednicki, M., 1981, Physics Letters B, 104, 199 
\bibitem{et}Etherington, I. M. H., 1933, Phil. Mag, 15, 761
\bibitem{et2}Etherington, I. M. H., 2007, Gen. Relativ. Gravit., 39, 1055
\bibitem{ellis2007}Ellis, G. F. R. 2007, Gen. Rel Grav., 39, 1047
\bibitem{Ellis13} Ellis, G.F.R., Poltis, R., Uzan, J.-P. Weltman, A., 2013, Phys. Rev. D, 87, 103530 
\bibitem{fuji}]  Fujii, Y \& Maeda, K. I., 2003, The Scalar-Tensor Theoryof Gravitation, by Yasunori Fujii and Kei-ichi Maeda, pp. 256. ISBN 0521811597. Cambridge, UK: Cambridge University Press
\bibitem{gon}Gonçalves, R. S.; Bernui, A.; Holanda, R. F. L.; Alcaniz, J. S., 2015, 573, A88
\bibitem{gon}Gon\c{c}alves, R. S., Holanda, R. F. L., \&  Alcaniz, J. S., 2012, MNRAS, 420, L43
\bibitem{gas}  Gasperini, M., Piazza, F. \& Veneziano, G., 2001, PRD,  65, 023508
\bibitem{guy}Guy, J., et al. 2007 A \& A, 466, 11
\bibitem{harko}Harko, T.,  Lobo, F. S. N., \& Minazzoli, O., 2013, PRD, 87, 047501
\bibitem{hees}  Hees, A., Minazzoli, O., \& Larena, J., 2014, PRD, 90, 124064
\bibitem{hoff}Hoff, P. D. (2009). \emph{ A first course in Bayesian statistical methods}. Springer Science \& Business Media.
\bibitem{h3a} Holanda, R. F. L., Lima, J. A. S. \& Ribeiro M. B., 2012b, A\&A, 538, 131
\bibitem{Holanda10} Holanda, R. F. L., Lima, J. A. S. \& Ribeiro, M. B., 2010, ApJ, 722, L233
\bibitem{Holanda} Holanda, R. F. L. \& Busti, V. C., 2014, PRD, 89, 103517
\bibitem{ha}Holanda, R. F. L., Gon\c{c}alves, R. S. \& Alcaniz, J. S., 2012a, JCAP, 06, 022
\bibitem{ho}Holanda, R. F. L., Carvalho, J. C. \&  Alcaniz, J. S., 2013, 04, 027
\bibitem{holanda3} Holanda, R. F. L., Busti, V. C. \& Alcaniz, J. S., 2016, JCAP accepted
\bibitem{hol1} Holanda, R. F. L.,  Landau, S. J.,  Alcaniz, J. S.,  Sanchez G., I. E. \& Busti,  V. C., 2016a, JCAP accepted, arXiv: 1510.07240
\bibitem{hol} R. F. L. Holanda, V. C. Busti, L. R. Colaço, J. S. Alcaniz \& S. J. Landau, 2016b, arXiv:1605.02578
\bibitem{ka} Kaplan, D. B., 1985, Nuclear Physics B 260, 215
\bibitem{ko}Kowalski, M., et al. 2008, ApJ, 749, 686
\bibitem{jaz} Lazkoz, R.,  Nesseris, S. \& Perivolaropoulos, L., 2008, JCAP, 0807, 012
\bibitem{liang} Liang, N. et al., 2013, MNRAS, 436, 1017
\bibitem{Liao} Liao, K. et al., 2015,  arXiv:1511.01318
\bibitem{liao2} Liao, K., Avgoustidis, A. \& Li, Z., 2015, PRD, 92, 123539 
\bibitem{Li}Li, Z., Wu, P. \& Yu, H., 2011, ApJL, 729, L14
\bibitem{mel}Melchiorri, F. \& Melchiorri, B. O., 2005, Proceedings of the International School of Physics “Enrico Fermi”, 159, 225
\bibitem{Meng12} Meng, X.-L., Zhang, T.-J., \& Zhan, H., 2012, ApJ, 745, 98
\bibitem{min} Minazzoli, O., Physics Letters B, 2014, 735, 119
\bibitem{min2}Minazzoli, O. \&  Hees, A., 2013, PRD 88, 041504
 \bibitem{Nair} Nair, R. Jhingan, S. \& Jain, D.,  2011, JCAP, 05, 023
\bibitem{over} Overduin, J. M., \& Wesson, P. S., 1997, Phys. Rep. 283, 303
\bibitem{pip} Peccei, R. D., \&  Quinn, H. R., 1977, PRD, 16, 1791
\bibitem{puxun} Puxun, W., Zhengxiang, L., Xiaoliang, L. \& Hongwei, Y., 2015, PRD, 92, id.023520
\bibitem{sand} Sandvik, H. B.,  Barrow, J. D., \& Magueijo, J., 2002, PRL, 88, 031302 
\bibitem{sanos}	Santos-da-Costa, S., Busti, V. C. \& Holanda, Rodrigo F. L., 2015, JCAP, 10, 061
\bibitem{saro} Saro, A., et al.  2014, MNRAS, 440, 2610 
\bibitem{sy}Sunyaev, R. A. \& Zel’dovich, Ya. B., 1972, Comments Astrophys. Space Phys., 4, 173
\bibitem{uzam}Uzan, J.P., Aghanim, N. and Mellier, Y., 2004, Phys. Rev. D, 70, 083533
\bibitem{yang} Yang, X. et al., 2013, JCAP, 06, 007
\bibitem{uzan}Uzan, J.P., Aghanim, N. \& Mellier, Y. 2004, PRD, 70, 083533
 \bibitem{whi}White, H., 1980, 817-838.
\end{thebibliography}
\end{document}